# Biological radiation dose from secondary particles in a Milky Way gamma ray burst

Dimitra Atri [1,2,*], Adrian L. Melott [3] and Andrew Karam [4]

1. Blue Marble Space Institute of Science, Seattle, WA 98145-1561, USA
2. Tata Institute of Fundamental Research, Colaba, Mumbai 400005, India
3. Department of Physics and Astronomy, University of Kansas, Lawrence, KS 66045, USA
4. Nevada Technical Associates, Inc., P.O. Box 93355, Las Vegas, NV 89193, USA

* Email: dimitra@bmsis.org

Abstract

Gamma ray bursts (GBRs) are a class of highly energetic explosions emitting radiation in a very short timescale of a few seconds and with a very narrow opening angle. Although, all GRBs observed so far are extragalactic in origin, there is a high probability of a GRB of galactic origin beaming towards the Earth in the past ~ 0.5 Gyr. We define the level of catastrophic damage to the biosphere as approximation 100 kJ per square meter, based on Thomas (2005 a,b). Using results in Melott and Thomas (2011), we estimate the probability of the Earth receiving this fluence from a gamma ray burst of any type, as 87% during the last 500 Myr. Such an intense burst of gamma rays would ionize the atmosphere and deplete the ozone layer. With depleted ozone, there will be an increased flux of solar UVB on the Earth's surface with potentially harmful biological effects. In addition to the atmospheric damage, secondary particles produced by gamma ray-induced showers will reach the surface. Amongst all secondary particles, muons dominate the ground-level secondary particle flux (99% of the total number of particles) and are potentially of biological significance. Using the Monte Carlo simulation code CORSIKA, we modeled the air showers produced by gamma ray primaries up to 100 GeV. We found that the number of muons produced by the electromagnetic component of hypothetical galactic GRBs significantly increases the total muon flux. However, since the muon production efficiency is extremely low for photon energies below 100 GeV, and because GRBs radiate strongly for only a very short time, we find that the biological radiation dose from secondary muons is negligible. The main mechanism of biological damage from GRBs is through solar UVB irradiation from the loss of ozone in the upper atmosphere.

1. Introduction

Ionizing radiation poses an intermittent threat to the Earth (Dartnell, 2011). Both supernovae and gamma-ray bursts (GRBs) may pose a threat at the level of mass extinction over geological time scales (Melott and Thomas, 2011; Horvath and Galante, 2012). Past consideration has focused on the danger from modification of atmospheric chemistry (Scalo and Wheeler, 2002; Melott et al., 2004). The GRB photons ionize and dissociate atmospheric constituents. The most important for this purpose is $N_2$, which is highly stable and tightly bound. The influx of free N atoms and ions into the atmosphere results in reactions that can produce various oxides of nitrogen, which in turn catalyze the destruction of stratospheric ozone ($O_3$). The loss of ozone in turn greatly reduces the opacity of the atmosphere to solar UVB radiation (Thomas et al., 2005a,b; Ejzak et al., 2007) – Cockell (1998) has calculated that, in the absence of an ozone layer, the DNA-weighted UV irradiance would be several hundred times higher than it is at present, depending on the age of the Sun (today's Sun is hotter and emits more UV than was the case when life first formed). Since UVB is strongly absorbed by DNA and protein, it can damage the exposed biota (both microorganisms and multicellular, complex organisms) and can cause erythema (skin reddening and inflammation such as sunburn), can cause cancer in complex organisms, can damage plankton and other organisms exposed to sunlight, and so forth. The end-Ordovician mass extinction has been proposed as a candidate for having a GRB as its initiator (Melott et al., 2004) and the extinction rate as a function of latitude for that event was found to be consistent with a burst over the south pole (Melott and Thomas, 2009).

Other dangers are possible, in particular secondary particles which will be created in the showers induced by high-energy photons. The purpose of this paper is to delineate the effect of these secondaries. This is an important check on past work, which has not included them.

Another approximation is that GRB jets have been assumed to be purely electromagnetic. Due to intervening magnetic fields, it is not possible to detect a hadronic component (likely protons and other atomic nuclei) from the large intergalactic distances of observed GRBs. However,, it is possible that significant emission of high-energy hadrons is possible and could strike the Earth if incident from within the Milky Way galaxy (e.g. Dermer and Holmes, 2005). However, the non-detection of neutrinos coincident with GRBs by the IceCube experiment has set limits on the hadron entrainment (Gao et al., 2013 and references therein). Since only upper limits exist, it is not possible to definitively model secondaries from hadrons. If there are any future detections, it will be possible to approach this problem with existing software and data tables (Atri and Melott, 2011).

2. Method

Since no two GRBs are identical, the choice of selecting a typical case of a galactic GRB becomes difficult. Since long GRBs, which last for ~10 s, have softer spectra, the photon energies will not be sufficient enough to produce muons. Short GRBs, also known as Short Hard Bursts (SHBs) on the other hand last for ~ 0.1 s and have harder spectra. A detailed review of a variety of observed GRBs with varying duration and spectra can be found in Band et al. (1993). The expected rate of SHBs affecting the Earth is slightly higher than due to long GRBs (Melott and Thomas 2011). They are expected to produce photons of sufficient energy to produce muons. We use two cases of hard-spectrum GRBs and evaluate their terrestrial effects. We used the Band et al. (1993) and GRB 120424A spectra (Augusto et al., 2013) for this work.

Based on the observation of a large number of GRBs, a phenomenological model was developed by Band et al. (1993), which provides a fit to the observed gamma ray spectra of a large number of GRBs. The energy of gamma-rays range from a few keV to ~1 GeV in most cases, and can go up to ~100 GeV in some rare extreme cases. For gamma-rays to undergo photonuclear interactions in the Earth's atmosphere and produce muons which can be observed on the ground, the energy of gamma-rays should be ~ 10 GeV and above. Muons can be produced by gamma rays of lesser energy, but since muons typically lose ~ 2-4 GeV energy through interactions (Gaisser, 1991) until they reach the ground, the energy threshold becomes higher, and only higher energy photons will produce > 2-4 GeV muons in substantial number. We will focus on GRBs that produce significant number of photons in the 10-100 GeV energy range, that have the capability of producing secondary particles reaching the Earth's surface.

2.1 Gamma ray spectra

Band spectrum: As discussed above, only short hard bursts are capable of producing GeV photons. Therefore, we took the Band spectrum (Band et al., 1993) for SHBs where the peak of the energy spectrum is highest, peaking at 187.5 MeV. It must be noted that SHBs are a subset of all GRBs and the peak of the energy spectrum is higher than usual. The spectral index in the 1-100 GeV range is -2.3. Our estimates are based on the fluence of photons received here, which is independent of the total energy/beaming angle interpretation. The spectrum was then normalized to obtain 100 kJ/$m^2$ fluence at the top of the Earth's atmosphere. The number of photons in 1-10 GeV and 10-100 GeV energy range were then computed based on the spectral index and fluence normalization. We obtained 6.29 x $10^{10}$ photons/$m^2$ in the 1-10 GeV range and 2.99 x $10^9$ photons/$m^2$ in the 10-100 GeV range for this spectrum after normalization.

GRB 120424A was observed with the MAXI/GSC instrument abroad the ISS where the prompt emission lasted for 34 seconds (Serino et al., 2012). Muons were observed with the ground-based Tupi muon observatory with a significance of 6.2 standard deviations (Augusto et al., 2013). This suggests muon production by gamma rays of energies greater than 10 GeV. Based on the muon data, the spectral index of -1.54 was found to be consistent for gamma ray primaries greater than 10 GeV.

2.2 Modeling gamma ray interactions in the atmosphere

GeV gamma rays interact with the Earth's atmosphere and produce air showers. Air showers were modeled with the state of the art CORSIKA model, a Monte Carlo code used extensively to model air showers and calibrated with experimental data with a variety of instruments worldwide. The code has already been shown to reproduce air shower data with great accuracy in some studies, including Usoskin et al., 2006; Atri et al., 2010; Atri and Melott, 2011; and Overholt et al., 2013. We ran CORSIKA simulations with -2.3 spectral index of primary gamma-rays to calculate the number of secondary muons at the surface per $10^7$ photon primaries in 1-10 GeV and 10-100 GeV energy ranges. The energy cutoff for muons was set to 50 MeV, which is the lowest possible option available in CORSIKA. It is clear from Figure 2 that such low energy muons were unable to reach the ground and therefore this energy cutoff does not affect our results.

3. Results

3.1 Particle flux

A total of $10^7$ gamma ray primaries with fixed energies were taken to be incident from random directions in the hemisphere away from the Earth, on top of the atmosphere. The energies are divided in logarithmic intervals in 10-100 GeV energy range. The hadronic interactions were modeled with the SIBYLL and FLUKA models. The total number of muons reaching the ground for each set of simulations is then calculated and displayed in form of a table (Table 1). As explained later in detail, there are total 0.0005% electrons and positrons, 0.000047% photons and no hadrons per primary reaching the ground. Therefore, only muons were taken into account in our calculations, and other particles were ignored.

Table 1: Total number of muons on ground for 10 million primaries in each case.

| Photon Energy (GeV) | Number of muons |
|---|---|
| 10 | 3535 |
| 12.5 | 5330 |
| 15.8 | 7512 |
| 19.9 | 9247 |
| 25.1 | 13401 |
| 31.6 | 18560 |
| 39.8 | 24300 |
| 50.1 | 32428 |
| 63.1 | 44025 |
| 79.4 | 58375 |
| 100 | 81057 |

The primaries in the computations were incident with random angles from the top of the atmosphere. A GRB is effectively a point source, and would only produce muon secondaries on one side of the Earth. The results presented here are effectively averages over the side of the Earth facing the GRB. In general, the muon flux falls as $\cos^2$ of the primary incident angle, measured from the vertical. The flux at the place where the GRB is exactly vertical will be about 1.6 times higher than the average presented here; at 45° it will be about 0.8 times these values.

Photon of energies less than 10 GeV were not simulated because the entire range produce less than 1% of muons produced in the 10-100 GeV range, as shown below.

For the Band spectrum (Band et al., 1993), 72 muons in the 1-10 GeV energy range and 10,406 muons in the 10-100 GeV range were produced from $10^7$ photons incident at the top of the atmosphere. In addition, in the 10-100 GeV range, 51 electrons and positrons and 47 photons were able to reach the ground from these $10^7$ primaries. There were no hadrons reaching the ground, as expected.

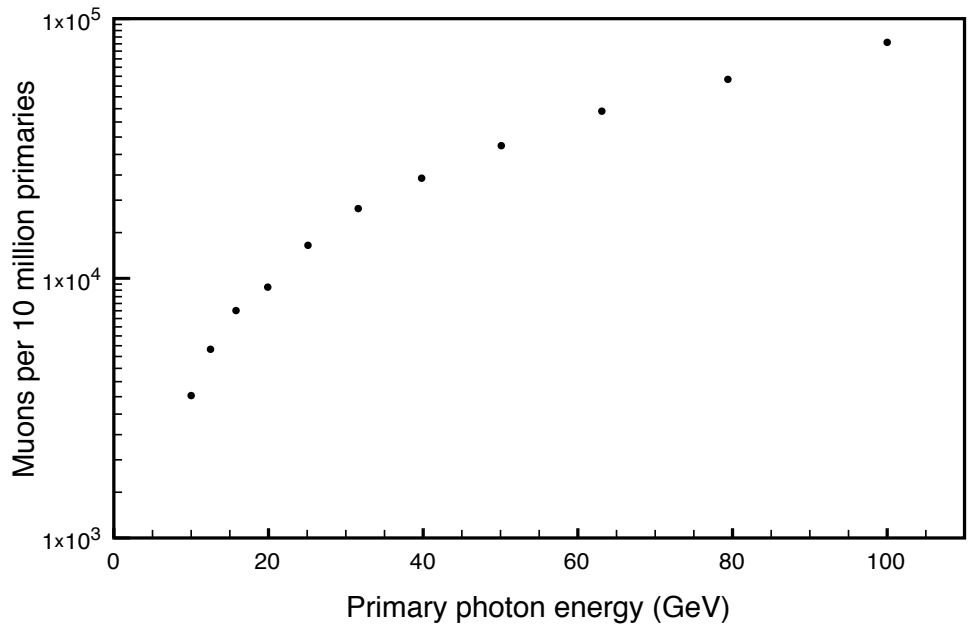

Figure 1: Total number of muons at the ground level per $10^7$ photon primaries as a function of photon energy

We adjust the total fluence of our GRB to 100 kJ/m$^2$, an amount likely to cause significant damage to terrestrial life every few hundred million years (Thomas et al. 2005b; Melott and Thomas 2011). For the normalized spectrum on top of the Earth's atmosphere at 100 kJ/m$^2$, we obtain at ground level 4.53x10$^5$ muons/m$^2$ in the 1-10 GeV and 3.12x10$^6$ muons/m$^2$ in the 10-100 GeV range, for a total of 3.57x10$^6$ muons/m$^2$ from the Band case, with a spectral index of -2.3. For comparison, the average rate of background cosmic ray muons integrated over the hemisphere is 127/m$^2$-sec. For a 1 sec burst, the muon flux enhancement would be by a factor of: (3.57x10$^6$)/127 = 2.81x10$^4$.

For GRB 120424A (Augusto et al., 2013), which is known to have produced muons on the ground, the reported observed rate was $1.78 \times 10^{-7}$ erg/cm$^2$/sec over 34 sec, which equals $6.05 \times 10^{-9}$ J/m$^2$ fluence. Using a spectral index of -1.54 from the above reference, we computed the muon fluence on the surface to be $6 \times 10^{-4}$ muons/m$^2$ at the ground integrated over 10-100 GeV range. We then normalized it for a total fluence of 100 kJ/m$^2$ (equivalent to moving it closer) and obtained $9.43 \times 10^9$ muons/m$^2$ in a period of 34 seconds. The muon rate therefore is $2.77 \times 10^8$/m$^2$/sec. The rate is about 2 orders of magnitude higher than the Band et al. case described above, due to the much harder spectrum and longer duration.

3.2 Biological radiation dose

Radiation dose is measured in units of Gray (Gy) and 1 Gy is equal to the deposition of 1 J of energy per kg of absorber. Different types of radiation are more effective at causing biological damage so 1 Gy of absorbed dose can cause different levels of biological damage depending on the density of energy transferred to the organism and the resulting density of damage to DNA and to other biologically important molecules; this is reflected by assigning a relative biological effectiveness (RBE) to various kinds of radiation where the RBE is related to the density of biological damage caused by the radiation. Alpha radiation, for example, causes roughly 20 times as much biological damage as does gamma radiation per unit of energy deposition; because of this alpha radiation is assigned an RBE of 20.

We should also note that the concept of RBE can be difficult to implement for microscopic organisms that are irradiated by radiation with a very short range in the organism. For example, alpha particles have a range of only 5-10 microns in material with the density of water – a single alpha particle can only traverse one or two cells. Consider a bacterial mat exposed to alpha radiation – any bacteria that are more than one or two cells deep will be spared any irradiation at all, in spite of a cell immediately above might receive a high (possibly lethal) radiation dose. Similarly, one cell on the surface of such a mat might receive a high dose of radiation while its immediate neighbors emerge unscathed. This can make alpha dosimetry very challenging. However, because muons are much more penetrating than are alpha particles, we feel it is safe to make the assumption that radiation dose from muons is uniform throughout an organism of any reasonable size. In this, we treat muon radiation as being similar to x-ray and gamma radiation in its impact on an organism and in its dosimetry. Although muons are very penetrating, they can have major biological effects due to their potential production in large numbers.

The amount of biological damage caused by radiation is measured in units of effective dose – Sievert (Sv. Effective dose is the product of absorbed dose (Gy) and the relative biological effectiveness of the radiation in question. Thus, the deposition of 1 J/kg of gamma radiation (RBE = 1) produces an effective dose of 1 Sv while depositing the same amount of energy via alpha radiation produces an effective dose of 20 Sv. The RBE of muons is about 0.8 – 1.0 across a wide range of energies (Stevenson, 1983; Micke et al. 1964; Nelson, 1966) so, for the sake of this work, the effective dose from muons (Sv) is approximately the same as the absorbed dose (Gy).

The biological impact of radiation exposure varies according to the dose and the organism exposed; higher doses are more damaging and simpler organisms (which typically have a smaller chromosome volume) tend to be more resistant to radiation damage than more complex organisms (UNSCEAR 1996). The time during which exposure occurs is also important and can also affect the biological impact; radiation exposure that occurs over a period of time that is shorter than the organism's ability to repair the induced biological damage (typically less than a day) will damage an organism while the same dose delivered over a longer period of time might leave an organism unscathed. In the case of exposure from a GRB the exposure occurs over a sufficiently short period of time that acute effects are possible if the radiation dose is sufficiently high.

The health effects of acute radiation exposure begin to occur at a dose of about 1 Sv (mild radiation sickness) for humans and the radiation exposure becomes fatal to humans at a dose of about 4-5 Sv, with 10 Sv being fatal to 100% of those exposed. Insects are known to be significantly more resistant to radiation than are mammals (the fatal dose for some insects is over 100 times as high as for humans), and microbes are more resistant still – *D. radiodurans* can survive exposure of up to 15 kSv of gamma or x-ray radiation. Other organisms fall between these extremes with regards to radiation resistance. Lower or more protracted radiation exposure might increase an organism's risk of developing cancer (in organisms sufficiently long-lived as to be susceptible to cancer) and can also induce germ-line mutations that might be able to affect rates of evolutionary change.

For this work a cube of water 15 cm on each side was considered as the target material for modeling the biological radiation dose from secondary muons. Figure 2 shows the energy spectrum of muons at the ground level. The muon energy spectrum peaks ~ 1 GeV and the flux sharply drops below 0.5 GeV. Since the stopping power of muons is a very slow function of energy, the average energy deposited by muons can be approximated by 2 MeV/g cm$^{-2}$ in this energy range (Gaisser, 1991). The biological radiation dose is the product of the muon flux within 225 cm$^2$ area, with 15 g/cm$^2$ column density with 2 MeV energy per gram and divided by 3.3375 kg of mass. This gives 1.14x10$^{-7}$ Sv or 0.11 $\mu$ Sv radiation dose from the Band case. Similarly, for GRB 120424A, the total biological radiation dose is 3.02x10$^{-4}$ Sv or 0.3 mSv. For comparison, the annual radiation dose limit for a terrestrial US worker is 50 mSv/yr.

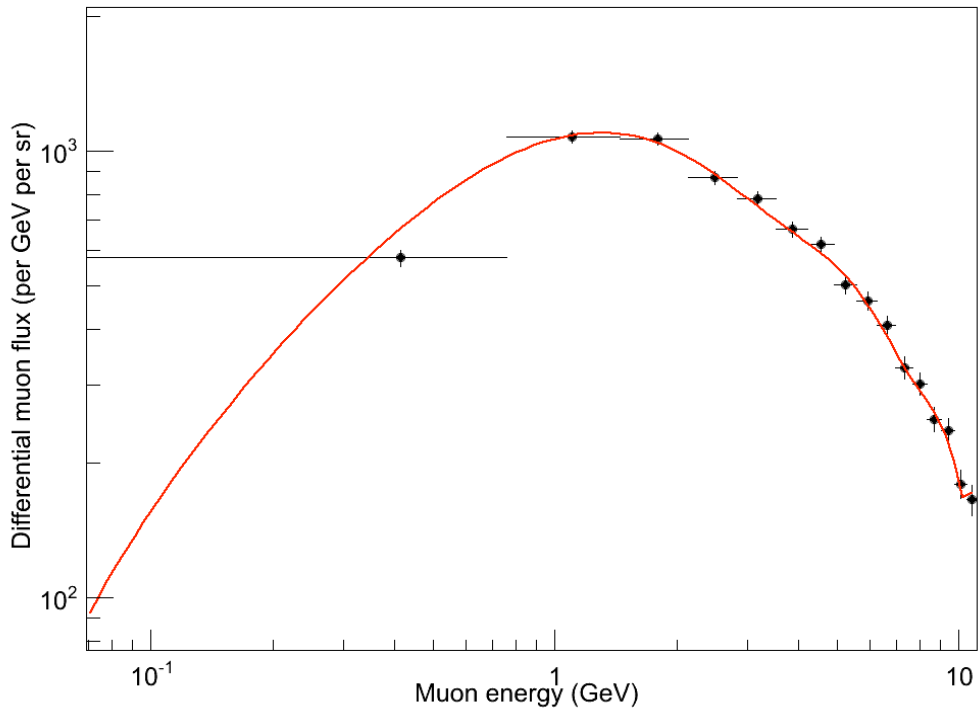

Figure 2: Differential spectrum of muons at the ground level obtained from CORSIKA for Band 1993 case.

4. Discussion

We have computed the expected fluence of secondaries in air showers generated by short-hard gamma ray bursts. Long bursts are typically too spectrally soft to generate secondaries by photonuclear reactions. We find that the irradiation by secondaries (primarily muons) is not a significant source of biological radiation dose. Consequently, past focus on atmospheric ozone depletion and subsequent UVB damage is appropriate. However, should future detection of neutrinos coincident with GRB bursts indicate significant hadronic component in GRB beams, this conclusion will require further evaluation.

5. Acknowledgments

We thank Brian Thomas for helpful comments on the manuscript.